\documentclass[a4paper]{article}

\usepackage{INTERSPEECH2019}
\usepackage{amsmath,graphicx,indentfirst,url,dirtree}
\usepackage{CJKutf8}

\newcommand{\Figure}[3]{\vspace{-0mm} \includegraphics[width=#1,clip]{#2} \vspace{-0mm} \caption{#3} \vspace{-0mm} \label{fig:#2}}

\newcommand{\drawfig}[4]{ 
  \begin{figure}[#1]
  \begin{center}
  \Figure{#2}{#3}{#4}
  \end{center} 
  \end{figure}
}

\title{JVS-MuSiC: Japanese multispeaker singing-voice corpus}

\name{Hiroki Tamaru, Shinnosuke Takamichi, Naoko Tanji, and Hiroshi Saruwatari}
\address{Graduate School of Information Science and Technology, The University of Tokyo, \\
7-3-1 Hongo Bunkyo-ku, Tokyo 133--8656, Japan.}

\email{shinnosuke\_takamichi@ipc.i.u-tokyo.ac.jp}

\begin{document}
\ninept
\maketitle
\begin{abstract}
Thanks to developments in machine learning techniques, it has become possible to synthesize high-quality singing voices of a single singer \cite{nishimura2016singing, blaauw2017neural}. An open multispeaker singing-voice corpus would further accelerate the research in singing-voice synthesis. However, conventional singing-voice corpora \cite{hts, jsutsong} only consist of the singing voices of a single singer. We designed a Japanese multispeaker singing-voice corpus called ``JVS-MuSiC'' with the aim to analyze and synthesize a variety of voices. The corpus consists of 100 singers' recordings of the same song, \textit{Katatsumuri}, which is a Japanese children's song. It also includes another song that is different for each singer. In this paper, we describe the design of the corpus and experimental analyses using JVS-MuSiC. We investigated the relationship between 1) the similarity of singing voices and perceptual oneness of unison singing voices and between 2) the similarity of singing voices and that of speech. The results suggest that 1) there is a positive and moderate correlation between singing-voice similarity and the oneness of unison and that 2) the correlation between singing-voice similarity and speech similarity is weak. This corpus is freely available online.
\end{abstract}


\section{Introduction}
Thanks to developments in machine learning techniques, it has become possible to synthesize high-quality singing voices of a single singer \cite{nishimura2016singing, blaauw2017neural}. An open multispeaker singing-voice corpus would further accelerate the research in singing-voice synthesis. However, conventional singing-voice corpora \cite{hts, jsutsong} only consist of the singing voices of a single singer. We designed a Japanese multispeaker singing-voice corpus called ``JVS-MuSiC'' with the aim to analyze and synthesize a variety of voices. The corpus consists of 100 singers' recordings of the same song, \textit{Katatsumuri}, which is a Japanese children's song. It also includes another song that is different for each singer. In this paper, we describe the design of the corpus and experimental analyses using JVS-MuSiC. We investigated the relationship between 1) the similarity of singing voices and perceptual oneness of unison singing voices and between 2) the similarity of singing voices and that of speech. The results suggest that 1) there is a positive and moderate correlation between singing-voice similarity and the oneness of unison and that 2) the correlation between singing-voice similarity and speech similarity is weak. This corpus is freely available online.

\section{Current Japanese singing-voice corpora} 
All Japanese singing-voice corpora consist of only a single singer's voices. The singing-voice corpus included in the demo of HTS \cite{hts} consists of 31 Japanese children's songs sung by a single female singer. JSUT-song \cite{jsutsong} consists of 25 songs in HTS sung by a different female singer. There is also a singing-voice database of synthesized singing voices called Tohoku Kiritan's singing-voice database \cite{kiritan}, which consists of 50 songs sung by Tohoku Kiritan, who is a female character of VOICEROID.

\section{Design of JVS-MuSiC}
\subsection{Structures}
The directory structures of the corpus are listed below. The singer name is formatted as \textit{jvs[SPKR\_ID]}, indicating the speaker ID with the range of 1 through 100.

The main purpose of JVS-MuSiC is to cover various singers' singing voices to enable the analysis and synthesis of the personality of singing voices. The 100 singers are the same as those of JVS corpus \cite{takamichi19jvs}; thus, it is also possible to investigate the relationship between speech and singing voices. The following sections describe how we designed the corpus.
    \dirtree{%
.1 \includegraphics[width=0.25cm]{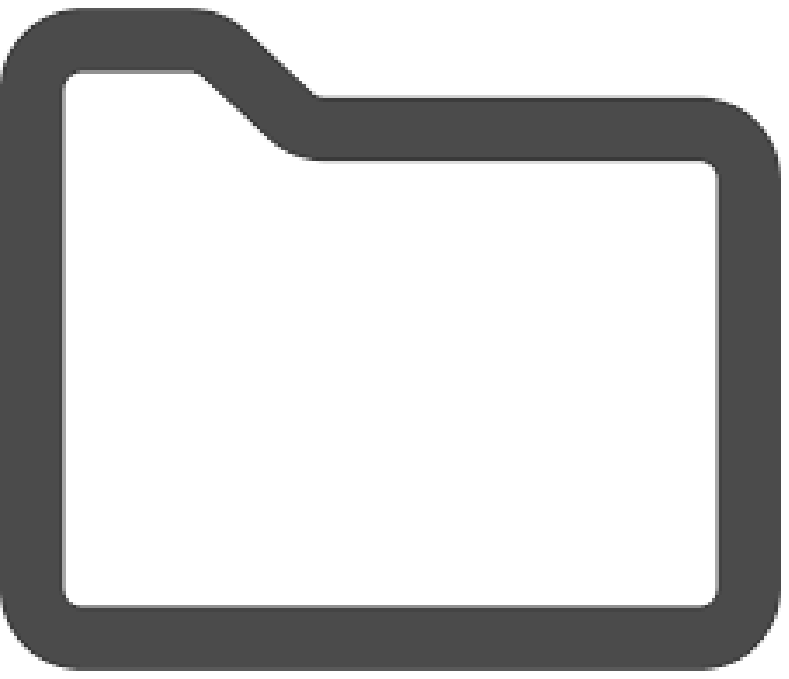}.
    .2 \includegraphics[width=0.25cm]{dir.eps} \textbf{jvs001}. 
        .3 \includegraphics[width=0.25cm]{dir.eps} \textbf{song\_common}.
            .4 \includegraphics[width=0.25cm]{dir.eps} \textbf{wav}.
                .5 \includegraphics[width=0.25cm]{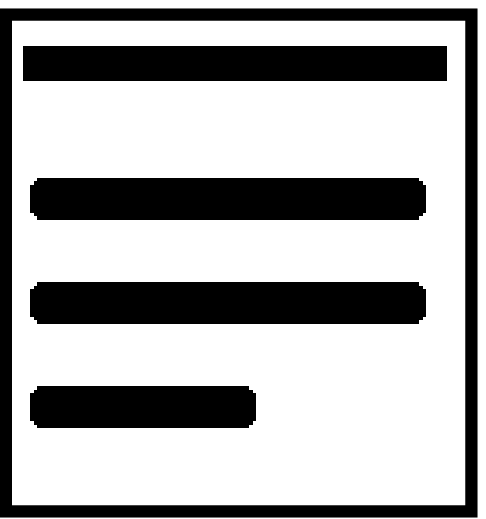} raw.wav.
                .5 \includegraphics[width=0.25cm]{file.eps} modified.wav.
                .5 \includegraphics[width=0.25cm]{file.eps} modified\_grouped.wav.
            .4 \includegraphics[width=0.25cm]{dir.eps} \textbf{mpd}.
                .5 \includegraphics[width=0.25cm]{file.eps} modified.mpd.
                .5 \includegraphics[width=0.25cm]{file.eps} modified\_grouped.mpd.
        .3 \includegraphics[width=0.25cm]{dir.eps} \textbf{song\_unique}.
            .4 \includegraphics[width=0.25cm]{dir.eps} \textbf{wav}.
                .5 \includegraphics[width=0.25cm]{file.eps} raw.wav.
    .2 \includegraphics[width=0.25cm]{dir.eps} \textbf{jvs002}.
    .2 \includegraphics[width=0.25cm]{dir.eps} ....
    .2 \includegraphics[width=0.25cm]{dir.eps} \textbf{jvs100}.
    .2 \includegraphics[width=0.25cm]{dir.eps} \textbf{similarity}.
        .3 \includegraphics[width=0.25cm]{file.eps} similarity\_\{name of group\}.csv.
    .2 \includegraphics[width=0.25cm]{dir.eps} \textbf{oneness}.
        .3 \includegraphics[width=0.25cm]{file.eps} oneness\_\{name of group\}.csv.
    .2 \includegraphics[width=0.25cm]{file.eps} singer\_info.txt. 
    }

\subsubsection{Raw voice (raw.wav)}
Recorded wav files of \textit{Katatsumuri} and singer-dependent songs are stored in song\_common/wav/ and song\_unique/wav/, respectively. Each singer sang \textit{Katatsumuri} in his or her favorite key and tempo; thus, the key and tempo vary among singers. The key and tempo are not completely consistent in each recording because the singers did not sing along with an example recording or a guide melody sound.

\subsubsection{Modified voices}
There are two versions of modified voices.
\drawfig{t}{1.0\linewidth}{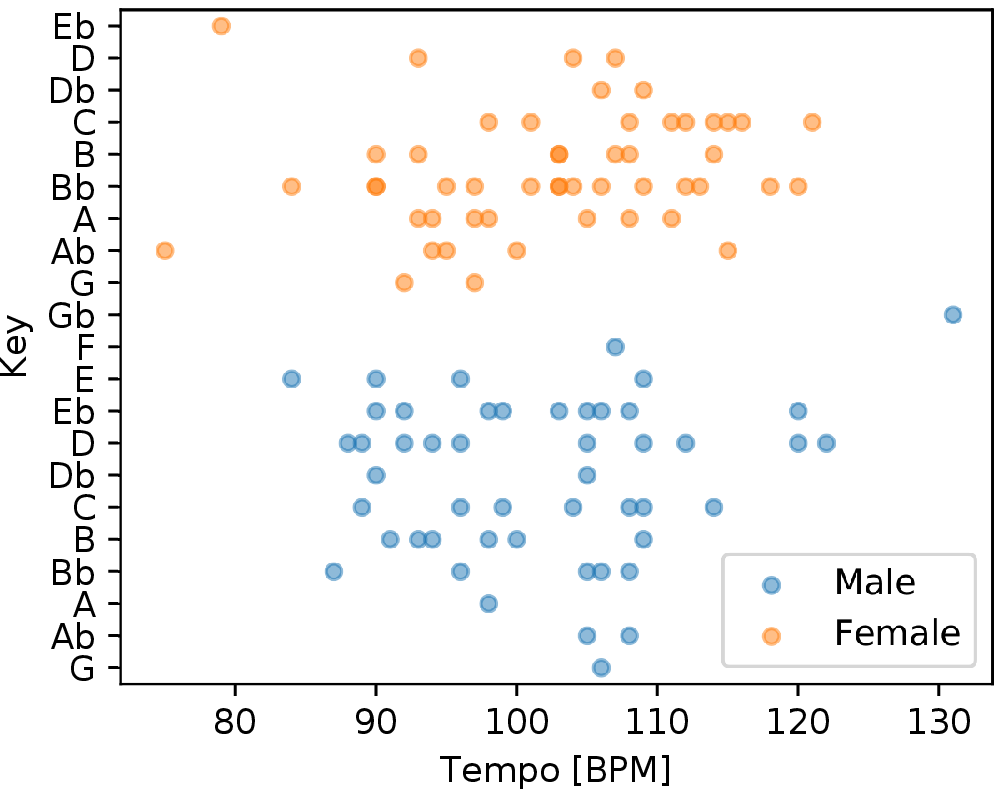} {Scatter plot of key and tempo.}
\begin{itemize}
    \item modified.wav\\
We determined the closest key and tempo for each singer and modified raw voices using the singing-voice modification software Melodyne \cite{melodyne}, as if the singer had sung in the key and tempo accurately. Figure~\ref{fig:scatter_tempo_key} shows the distribution of the keys and tempos of modified voices. There is a four-beat silence at the beginning of each file. The key and tempo labels of the modified voices are stored in mpd\_label.txt, which consists of speaker ID, gender, tempo (BPM: beats per minute), and key number (how many semitones higher than the lowest one). 
    \item modified\_grouped.wav\\
For facilitating analysis, we created a grouped version of modified wav files and divided the 100 singers into six groups (three for each gender) by key. We used Melodyne to unify the key within the group and the tempo within all groups (100 BPM). We chose Melodyne for this purpose because it produces empirically less change in voice timbre than with a time stretching- and resampling-based method. Table \ref{tab:group} lists the keys and numbers of singers for the groups. 

\begin{table}[t]
\centering
\caption{Description of singer groups}
\begin{tabular}{c|cc} \hline \hline
Group & Key & \# of singers \\ \hline
Male-low & B$\flat$ & 17 \\
Male-middle & D$\flat$ & 16 \\
Male-high & E$\flat$ & 16 \\
Female-low & A & 17 \\
Female-middle & B$\flat$ & 17\\
Female-high & D$\flat$ & 17 \\
\hline
\end{tabular}
\label{tab:group}
\end{table}

\end{itemize}

\subsubsection{MPD file (*.mpd)}
modified.mpd and modified\_grouped.mpd are Melodyne project documents that were used to create the corresponding modified voices. We used Melodyne 4 Assistant \cite{melodyne} to modify singing voices. The editor was not a professional engineer, but had some experience in music production including vocal editing as an amateur. We now describe the modification steps.

\begin{enumerate}
\item Determination of key and tempo

We determined the closest key and tempo for each singing voice by listening to the voice and looking at the graphical user interface.

\item Editing pitch

We used the correct pitch macro to modify the pitch of musical notes. We set both pitch center and pitch drift to 100\% then manually checked and modified all notes considering the perceptual naturalness.

\item Editing time

We used the time tool to manually modify the onset and offset of all musical notes considering the perceptual naturalness.
\end{enumerate}

\subsubsection{Similarity and oneness matrices (*.csv)}
These csv files are the similarity and oneness matrices of the experimental analysis mentioned in Section \ref{exp}.

\subsection{Recording}

We hired 100 native Japanese professional speakers; 49 males and 51 females. Their voices were recorded in a recording studio, and the recording for each speaker was done within one day. The recordings were controlled by a professional sound director. The voices were originally sampled at 48~kHz and downsampled to 24~kHz by the Speech Signal Processing Toolkit \cite{sptk}, and the 16-bit/sample RIFF WAV format was used. These settings are the same as those in the JVS corpus~\cite{takamichi19jvs}. The total duration of the 100 files of the common song was 49 min and 23 sec, and that of singer-dependent songs was 88 min and 3 sec.

\section{Experimental analysis}
\label{exp}
\subsection{Experimental conditions}
We evaluated the inter-speaker similarity and oneness of unison for many pairs of singers. We used grouped files because the key should be unified when annotating similarity and must be unified when producing unison voices and because too much pitch shifting may cause artifacts and changes in voice timbre. For annotating perceptual similarity scores, we used 9.6-sec samples, which were made by concatenating two singers' voice samples of the same musical phrase of 4.8~sec (eight beats). We followed Saito et al.'s study \cite{saito19perceptual}in which each listener scored the perceptual similarity for each pair of speakers from $-3$ (completely different) to $+3$ (very similar). For the evaluation of the oneness of unison, we used 9.6-sec (16 beats) samples of two singers' unison singing-voice samples. The unison voices were obtained by mixing two singers' voices. We balanced the volume of the two voices by equalizing the mean squared amplitude when mixing. The listeners scored the separateness of unison (i.e., how much the two singers' voices were heard separately, not as a united one) from 1 (heard as one) to 5 (heard separately). The oneness of unison was obtained by inverting the sign of separateness of unison. A final score for each speaker pair was obtained by averaging listeners' scores. For analysis, we normalized the measured values into the range (0, 1), used the crowdsourcing platform ``Lancers'' \cite{lancers}, and gathered ten participants for each pair of singers.

\subsection{Correlation between similarity and oneness of unison}

Figure~\ref{fig:scatter_sim_uni_oneness_1227} is the scatter plot of average similarity and oneness of unison for all 784 ($=4{\binom{17}{2}}+2{\binom{16}{2}}$) pairs. The correlation coefficient is 0.45 and $p$-value is $1.4\times 10^{-39}$. Table \ref{tab:r_and_p} shows the group-wise results, which suggest that, to some extent, a pair of singers with similar voices produces a united unison voice, which is often considered to sound beautiful.

\drawfig{t}{1.0\linewidth}{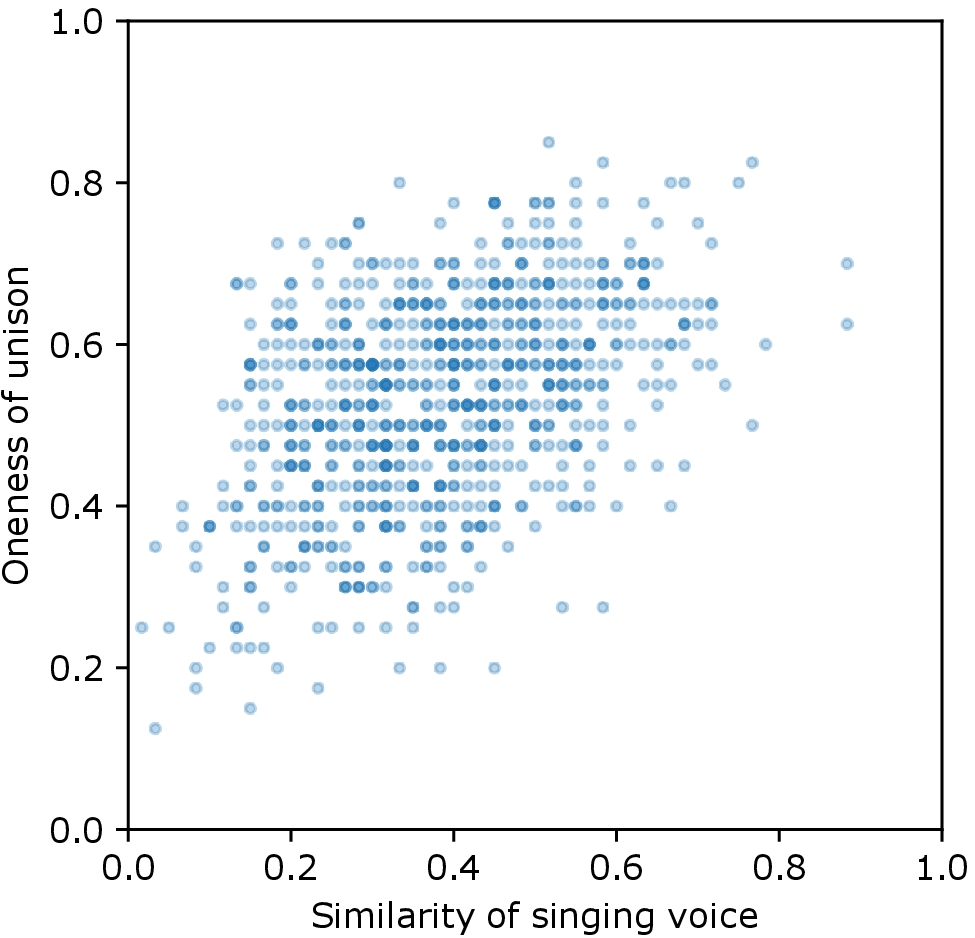} {Scatter plot of average similarity and oneness of unison. Correlation coefficient is 0.45 and $p$-value is $1.4\times 10^{-39}$.}

\begin{table}[t]
\centering
\caption{Correlation coefficients ($r$) and $p$-values for all groups}
\begin{tabular}{c|cc} \hline \hline
Group & $r$ & $p$-value \\ \hline
Male-low & $0.28$ & $8.2\times 10^{-4}$ \\
Male-middle & $0.58$ & $<10^{-10}$ \\
Male-high & $0.53$ & $3.6\times 10^{-10}$ \\
Female-low & $0.30$ & $3.9\times 10^{-4}$ \\
Female-middle & $0.36$ & $1.9\times 10^{-5}$ \\
Female-high & $0.51$ & $2.0\times 10^{-10}$ \\
All & $0.45$ & $<10^{-10}$ \\
\hline
\end{tabular}
\label{tab:r_and_p}
\end{table}

\subsection{Correlation between singing-voice similarity and speech similarity}

\drawfig{t}{1.0\linewidth}{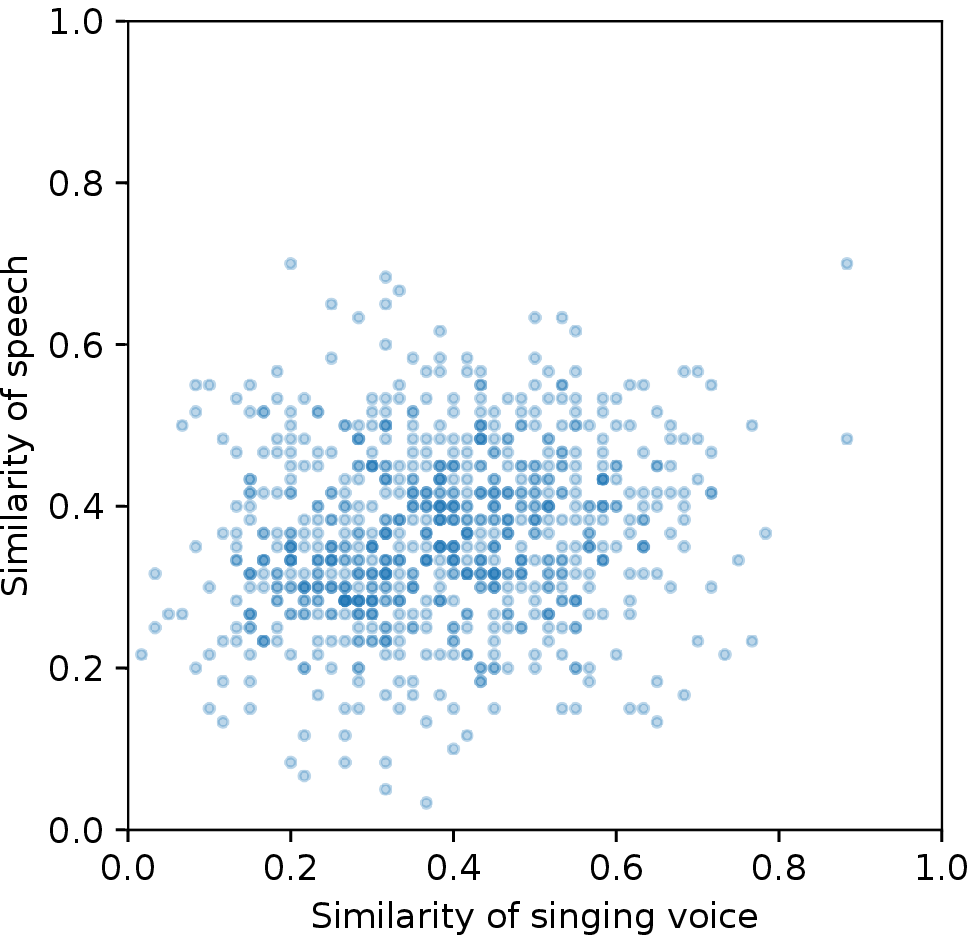} {Scatter plot of average similarity of singing voice and speech. Correlation coefficient is 0.17 and $p$-value is $1.9\times 10^{-6}$.}

Figure~\ref{fig:scatter_sing_speech} is the scatter plot of the average similarity of singing voice and similarity of speech \cite{takamichi19jvs} for the 784 pairs. The correlation coefficient is 0.17 and $p$-value is $1.9\times 10^{-6}$. Although part of this low correlation may be due to the change in voice timbre caused by pitch shifting, this result suggests that a person's singing voice and speech are fundamentally different or that listeners perceive singing voices and speech in different manners.

\section{Conclusion}
We introduced a Japanese multispeaker singing-voice corpus called JVS-MuSiC. This corpus was designed for multispeaker singing-voice analysis and synthesis and can be used for research in unison singing voices. We analyzed the relationship between the similarity of singer pairs and unison singing voices. The experimental results suggest that there is a positive and moderate correlation between the similarity and oneness of unison. We also compared singing-voice similarity to speech similarity and found that the correlation between them was weak.

The similarity and oneness matrices are licensed with CC BY-SA 4.0. The audio data and MPD files may be used for 
\begin{itemize} \itemsep 0mm
\item Research by academic institutions
\item Non-commercial research, including research conducted within commercial organizations
\item Personal use, including blog posts.
\end{itemize}
Our project page at \url{https://sites.google.com/site/shinnosuketakamichi/research-topics/jvs_music} describes the terms for commercial use.

\textbf{Acknowledgments:} Part of this work was supported by the SECOM Science and Technology Foundation and the GAP foundation program of the University of Tokyo.

\ninept
\bibliographystyle{IEEEbib}
\bibliography{tts}

\begin{thebibliography}{10}

\bibitem{nishimura2016singing}
M.~Nishimura, K.~Hashimoto, K.~Oura, Y.~Nankaku, and K.~Tokuda,
\newblock ``Singing voice synthesis based on deep neural networks,''
\newblock in {\em Proc. INTERSPEECH}, San Francisco, U.S.A., Sep. 2016, pp.
  2478--2482.

\bibitem{blaauw2017neural}
M.~Blaauw and J.~Bonada,
\newblock ``A neural parametric singing synthesizer modeling timbre and
  expression from natural songs,''
\newblock {\em Applied Sciences}, vol. 7, no. 12, Dec. 2017.

\bibitem{hts}
``{HMM}-based speech synthesis system ({HTS}),'' http://hts.sp.nitech.ac.jp/.

\bibitem{jsutsong}
``{JSUT}-song,'' \url{https://sites.google.com/site/shinnosuketakami
  chi/publication/jsut-song}.

\bibitem{kiritan}
``Kiritan singing database,'' \url{https://zunko.jp/kiridev/login.php}.

\bibitem{takamichi19jvs}
S.~Takamichi, K.~Mitsui, Y.~Saito, T.~Koriyama, N.~Tanji, and H.~Saruwatari,
\newblock ``{JVS} corpus: free {J}apanese multi-speaker voice corpus,''
\newblock {\em arXiv}, vol. abs/1908.06248, 2019.

\bibitem{melodyne}
``Celemony | what is melodyne?,''
  https://www.celemony.com/en/melodyne/what-is-melodyne.

\bibitem{sptk}
``Speech signal processing toolkit ({SPTK}) http://sp-tk.sourceforge.net/,''
\newblock .

\bibitem{saito19perceptual}
Y.~Saito, S.~Takamichi, and H.~Saruwatari,
\newblock ``{DNN}-based speaker embedding using subjective inter-speaker
  similarity for multi-speaker modeling in speech synthesis,''
\newblock in {\em Proc. SSW10}. Vienna, Austria, Sep. 2019.

\bibitem{lancers}
``Lancers,'' \url{https://www.lancers.jp/}.

\end{thebibliography}

\end{document}